\documentclass[sigconf]{acmart}

\AtBeginDocument{%
  }
\setcopyright{none}
\renewcommand\footnotetextcopyrightpermission[1]{}
\usepackage{booktabs}
\usepackage{multirow}
\usepackage{longtable}
\usepackage{xspace}
\usepackage{enumitem}

\usepackage{pseudocode}
\usepackage[normalem]{ulem}
\usepackage{caption}
\usepackage{subcaption}
\usepackage{tikz}
\usepackage{amsmath}
\usepackage{graphicx}
\usepackage{epstopdf}
\usepackage{listings}
\usepackage{hyperref}
\usepackage{url}
\usepackage{amsthm}

\usepackage[utf8]{inputenc}
\usepackage[english]{babel}

\theoremstyle{definition}
\newtheorem{definition}{Definition}

\lstdefinelanguage
[x64]{Assembler}     
[x86masm]{Assembler} 
{morekeywords={CDQE,CQO,CMPSQ,CMPXCHG16B,JRCXZ,LODSQ,MOVSXD, %
		POPFQ,PUSHFQ,SCASQ,STOSQ,IRETQ,RDTSCP,SWAPGS, addl, addq, cltq, %
		movslq, movzbl, %
		rax,rdx,rcx,rbx,rsi,rdi,rsp,rbp, %
		r8,r8d,r8w,r8b,r9,r9d,r9w,r9b, %
		r10,r10d,r10w,r10b,r11,r11d,r11w,r11b, %
		r12,r12d,r12w,r12b,r13,r13d,r13w,r13b, %
		r14,r14d,r14w,r14b,r15,r15d,r15w,r15b}} 

\newcommand{\sys}{VarMRI\xspace}
\newcommand{\utask}{\emph{AppTask}\xspace}
\newcommand{\utasks}{\emph{AppTask}s\xspace}
\newcommand*{\myfont}{\fontfamily{cmtt}\smaller\selectfont}

\DeclareTextFontCommand{\code}{\myfont}

\def\TR{}

\begin{document}

\title{Long-term Monitoring of Kernel and Hardware Events to Understand Latency Variance}

	\author{Fang Zhou}
	\affiliation{%
		\institution{The Ohio State University}
		\streetaddress{}
		\city{}
		\state{}
		\country{}}

	\author{Yuyang Huang}
	\affiliation{%
		\institution{The Ohio State University}
		\streetaddress{}
		\city{}
		\state{}
		\country{}}
	
	\author{Miao Yu}
	\affiliation{%
		\institution{The Ohio State University}
		\streetaddress{}
		\city{}
		\state{}
		\country{}}
	
	\author{Sixiang Ma}
	\affiliation{%
		\institution{The Ohio State University}
		\streetaddress{}
		\city{}
		\state{}
		\country{}}
		
     	\author{Tongping Liu}
	\affiliation{%
		\institution{University of Massachusetts Amherst}
		\streetaddress{}
		\city{}
		\state{}
		\country{}}
	
	\author{Yang Wang}
	\affiliation{%
		\institution{The Ohio State University}
		\streetaddress{}
		\city{}
		\state{}
		\country{}}


\begin{abstract}
\vspace{.05in}

This paper presents our experience to understand latency variance
caused by kernel and hardware events, which are often invisible at the
application level. For this purpose,
we have built \sys, a tool chain to monitor and analyze those events in the long
term. To mitigate the ``big data'' problem caused by long-term monitoring, \sys
selectively records a subset of events following two principles: it only
records events that are affecting the requests recorded
by the application; it records coarse-grained information first and records
additional information only when necessary. Furthermore, \sys introduces
an analysis method that is efficient on large amount of data, robust on different
data set and against missing data, and informative to the user.

\sys has helped us to carry out a 3,000-hour study of six applications and
benchmarks on CloudLab.
It reveals a wide variety of events causing latency
variance, including interrupt preemption, Java GC, pipeline stall, NUMA balancing
etc.; simple optimization or tuning can reduce tail latencies by up to 31\%. Furthermore,
the impacts of some of these events vary significantly across different experiments,
which confirms the necessity of long-term monitoring.

\end{abstract}

\maketitle

\section{Introduction}

This paper presents our experience to understand latency variance caused
by kernel and hardware events. It includes \sys, a tool chain we have developed for long-term monitoring and analysis,
and a 3000-hour study of six applications and benchmarks on CloudLab~\cite{cloudlab} with the help of \sys.

Latency is one of the most important performance metrics of server applications.
Latencies of processing even the same type of requests can vary a lot because
of contention, slow hardware, background 
activities, etc.~\cite{Maricq2018Variability,Kanev15Profiling,Gunawi18Failslow,GanASPLOS2019,PanoramaOSDI2018,Dean2013Tail,Gunawi18Failslow,DelimitrouCACM2018}.
Such variance motivates service providers to optimize their tail latencies
to satisfy most of their users~\cite{Dean2013Tail,Decandia07Dynamo}.
A good understanding of the causes of latency variance
can often help developers to mitigate the problem by re-designing the system~\cite{Decandia07Dynamo,Berger2018RobinHood,Qin2018Arachne,Sriraman2018uTune,Fried2020Caladan},
tuning configuration parameters~\cite{Li2014Tales,WangASPLOS2018,AshrafATC2019,ZhuSOCC2017}, or changing hardware~\cite{Maricq2018Variability,WangDSN2017,Zhou2018wPerf}.

The unpredictable nature of latency variance requires long-term monitoring
to understand their causes. Indeed, industrial
companies have been building such monitoring tools, such as Google Dapper~\cite{Sigelman2010Dapper},
Facebook Canopy~\cite{Kaldor2017Canopy}, and Twitter Zipkin~\cite{zipkin}. However,
these tools mainly record events in the application and thus cannot analyze the impacts of kernel and hardware events
like thread scheduling, cache misses, etc. This is not ideal since these
events are known to have a significant impact on tail latency~\cite{Li2014Tales,Qin2018Arachne,Sriraman2018uTune,Fried2020Caladan}.

Given that today's OS and hardware have already provided support for recording kernel and hardware events~\cite{ftrace,kprobe,ebpf,dtrace,etw},
adding them into latency analysis look like a straightforward task,
but our experiments encounter the classic challenge of ``big data'': if we naively record all
events, a single Memcached server can generate 104MB of trace per second
(i.e., 9TB of trace per day). A user is unlikely to provide the additional storage space
and I/O bandwidth to store such large amount of traces; even if she did, analyzing
the trace will consume much resource.

Such results indicate that we can only afford to record a
small subset of events, which is a standard solution to reduce overhead, but as
a result, we need to answer the
key question \emph{how to select what events to record to
minimize the recording overhead while still allowing meaningful analysis.}
Answering this question requires a collaborative design of trace recording
and trace analysis.

\vspace{.1in}
\noindent{\textbf{Efficient online recording.}}
To minimize overhead without disrupting the following analysis, \sys's event recording follows two principles: first, given that
latency analysis is performed at the granularity of application requests,
if the application decides to
record a subset of requests to limit overhead~\cite{Sigelman2010Dapper,Kaldor2017Canopy}, \sys should only record kernel
and hardware events that are affecting these selected requests.
We find that, among the different recording granularities provided
by existing tools (i.e., the whole
machine, a process, a thread, or a CPU core), recording
events at thread granularity best fits with this principle, since thread is the basic unit to 
execute requests in multi-threaded applications.

Our second principle is to record coarse-grained information first to minimize
overhead and to record additional information only when necessary.
We find that, among the different
recording modes provided by existing
tools (i.e., timing, call stack, or cumulative value), recording 
cumulative values of events incurs the least I/O overhead per event and can satisfy most of our goals,
so \sys records cumulative values of kernel and hardware events first,
and then records their exact timing or call stacks only when necessary.

While existing tools have implemented some of the features discussed above,
our study also reveals important functions that are missing, including recording the cumulative
lengths of events and separating the effects of interrupts from thread execution. \sys adds these missing
pieces. Among them, we find recording cumulative length is challenging
because kernel events can preempt each other and thus
it is hard to compute their precise lengths. To address
this problem, \sys maintains a per-core \emph{shadow stack} to simulate how the OS kernel
keeps track of the preemption relationship, which can be updated efficiently at run time.

\vspace{.1in}
\noindent{\textbf{Efficient, informative, and robust offline analysis.}}
\sys's offline analysis needs to find events that are causing long latency.
It should satisfy several key properties:
1) efficiency: 
analysis with a 
complexity of $O(N^2)$ (N is the number of recorded requests) or more
is not preferred; 2) informative: the analysis should tell the developer
which events are more worth optimizing, which is complicated by the
facts that many events are correlated;
3) robustness: the analysis should minimize its mathematical assumptions on the
trace data and should work with ``missing data'' since selective recording
only records a subset of events for each request.

Our attempts
with classic statistical methods have all failed due to violating at least one
of these properties (see Section~\ref{sec:statistics} for details),
which motivates us to develop new methods. To quantify the impact
of an event, we introduce a novel metric called \emph{impact value}, which estimates
the benefit of optimizing an event by comparing request latencies with and without high
values of the event: this metric is informative to the developer, easy to compute,
and does not rely on specific mathematical assumptions.  We further develop two methods
to facilitate the computation of the impact value: first, \sys applies line-based curve fitting on the CDF of an event's values 
to identify ``high'' value of the event; 
second,
\sys incorporates two domain-specific
rules to infer the causal relationship among events and adjust their impact values
accordingly.

\vspace{.1in}
\noindent{\textbf{Results from long-term experiments.}
With the help of \sys,
we have carried out a 3000-hour study of six applications and benchmarks
on CloudLab~\cite{cloudlab}.

\begin{itemize}[leftmargin=*]

\item \sys has identified a variety of kernel and hardware events causing
tail latencies within each experiment;
optimizing them can reduce tail latencies by up to 31\%, despite that we have
avoided complicated optimizations. The prevalence
and impact of these events have confirmed the importance of including them
in latency variance analysis.

\item \sys has also identified events whose impacts can vary across different
experiments, which can explain the performance variance. For example,
we find the burstiness of interrupts can vary: more bursty interrupts can cause higher tail latency
but lower median latency.
Such results have confirmed the necessity of long-term monitoring.

\item \sys's support of selective recording allows one to adjust the overhead by tuning the
selection rate. For example, recording events for 1\% of the requests in Memcached generates
1MB of trace per second, reduces maximal throughput by 0.49\%, and has no
observable impact on p99 latency. 

\end{itemize}

	\section{Background and related work}
\label{sec:related}

Many previous works have discussed the importance and the possible causes of latency variance~\cite{Dean2013Tail,Decandia07Dynamo,Li2014Tales,Arbel2018BST,Pesterev2010Cache,Maricq2018Variability,Jung2014HIOS},
which have motivated a number of analysis tools~\cite{Huang2017VProfiler,Barham2004Magpie,Attariyan2012XRay,Chen2004PPF,Fonseca2007Xtrace,Sigelman2010Dapper,Reynolds2006Pip,Mace2015Retro,Mace2015Pivot,Sambasivan2011DPC,Chow2014Mystery,Kaldor2017Canopy,Zhao2014lprof,Zhao2016Stitch,Rogora2020PerfAnno}.
They can record and analyze different types of events in the following ways:

\vspace{.1in}
\noindent{\textbf{Application events.}
At the application level, the standard solution is to annotate application code
to print the timing of different code blocks (called \utasks in this paper): by comparing
the patterns of different requests, we may find code blocks that are only
executed or take longer to execute in long-latency requests.
Existing tools can identify important code blocks by relying on developers'
annotations~\cite{Barham2004Magpie,Sigelman2010Dapper,Chow2014Mystery,Kaldor2017Canopy,Chen2004PPF,Reynolds2006Pip},
extracting events from the application logs~\cite{Zhao2014lprof,Xu2009DLS}, identifying events automatically~\cite{Attariyan2012XRay},
or using a mixture of these techniques~\cite{Huang2017VProfiler,Sigelman2010Dapper}.

In a multi-threaded or distributed application, the execution of a request may create multiple \utasks spanning different
threads or processes. To find the critical path of such complicated
execution, existing tools can model request execution using paths~\cite{Chen2004PPF},
trees~\cite{Fonseca2007Xtrace,Sigelman2010Dapper,Chanda2007Whodunit,Barham2004Magpie},
or directed acyclic graphs (DAG)~\cite{Reynolds2006Pip,Mace2015Pivot,Sambasivan2011DPC,Fonseca2010ETC,Mace2015Retro}.

\vspace{.1in}
\noindent{\textbf{Kernel events.}
Modern OSes have built-in tools to record kernel events (e.g., \emph{perf}, \emph{eBPF},
\emph{ftrace}, \emph{kprobe}/\emph{kretprobe} in Linux~\cite{ftrace,kprobe,ebpf,perf}; DTrace in Solaris~\cite{dtrace}; ETW in Windows~\cite{etw}, etc.),
which have been continuously improved~\cite{Yaghmour2000,Erlingsson2011Fay}.
These OSes have predefined many events, called \emph{trace points},
and these tools can record these predefined events in different modes: they can record
events on the whole machine, on a particular CPU core, on a particular process, or on a particular
thread; they can record the timing of these events, the call stack
of them, or the cumulative count of them.
A few of the analysis tools mentioned above, such as DARC~\cite{Traeger2008DARC}  and Magpie~\cite{Barham2004Magpie},
rely on OS built-in tools to record kernel events. However, as shown in Section~\ref{sec:kernel}, we encounter
additional challenges when utilizing existing OS built-in tools for
long-term monitoring.

\vspace{.1in}
\noindent{\textbf{Hardware events.}
Many hardware devices provide performance counters to help developers to investigate their internal behavior.
To limit overhead, some devices can only record a small subset
of performance counters at a time. For example, Intel Xeon Silver 4114 CPU provides
201 performance counters but can only record seven of them
at a time (three of them are fixed and four of them are configurable)~\cite{intelmsr}.

	\begin{figure}[t]

\centering
 \includegraphics[width=0.45\textwidth]{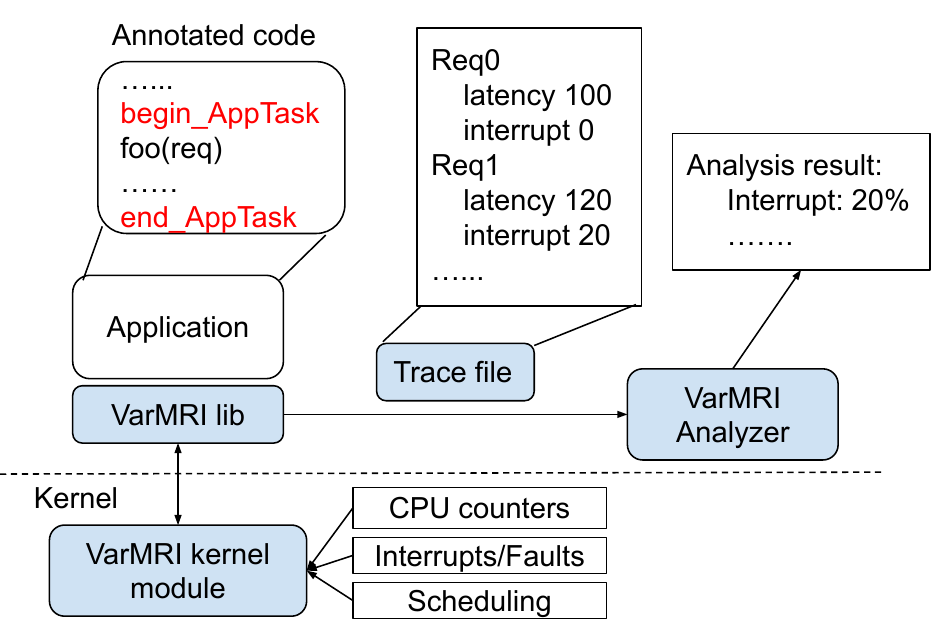}
 \vspace{-.1in}
 \caption{The architecture of \sys.}
 \label{fig:architecture}
 \vspace{-.15in}
\end{figure}

\section{Overview of \sys}
\label{sec:overview}

\sys records kernel and hardware events at run time
and analyzes their impacts on request latency offline.
\sys does not target events in the application,
so for a comprehensive analysis, it should be used together with
existing application-level tools like Zipkin.

\vspace{.1in}
\noindent{\textbf{Kernel events.}
\sys records information about thread scheduling, interrupts, and
faults because they are usually invisible to the application ~\cite{Traeger2008DARC,Zhou2018wPerf}.
\sys does not record system calls because they can be captured
by the application.

\vspace{.1in}
\noindent{\textbf{Hardware events.}
\sys records performance counters from
the CPU, such as the number of cache
misses. 
In addition, \sys computes the average CPU frequency (i.e., $\frac{cycles}{time}$)
as an additional counter.
Although the current implementation of \sys does not
record events from other types of hardware, it can easily be extended
to support hardware which provides similar performance counters.

\vspace{.1in}
\noindent{\textbf{Architecture and workflow of \sys.}
Figure~\ref{fig:architecture} shows the architecture of \sys.
A developer should first annotate the
target piece of code to analyze with the APIs provided by \sys: this can be done
either manually or automatically with the help of other tools~\cite{Attariyan2012XRay,Huang2017VProfiler,Traeger2008DARC}, and we observe
many softwares already annotate and record the execution of important
pieces of code. Then the developer can run the annotated application
with \sys's online modules (i.e., \sys lib and \sys kernel module; Section~\ref{sec:kernel}), which
will record the values of kernel and hardware events into a trace file.
Finally she should run the \sys analyzer offline to generate a list
of events ranked by their impacts (Section~\ref{sec:statistics}).
This architecture is not much different from 
other tools~\cite{perf,oprofile,ftrace,Barham2004Magpie,Traeger2008DARC}.

\vspace{.1in}
\noindent{\textbf{Limitation of \sys.}
Like most performance profiling tools~\cite{perf,oprofile,yourkit,gprof,ftrace}, \sys 
only reports potential problems to the developer, but 
does not address these problems.

\sys selectively records a subset of events to limit overhead.
An inherent limitation of selective recording is that it may miss very rare
events, if they happen to be not selected when they occur. Therefore, \sys is only
effective when problematic
events are reproducible with a non-negligible possibility. However, given
that a number of production
systems (e.g., Google Dapper~\cite{Benjamin2010Dapper}, Facebook Canopy~\cite{Kaldor2017Canopy}, etc.)
already adopt selective recording at the application level to reduce overhead, this limitation is arguably inevitable.

\vspace{.1in}
\noindent{\textbf{Motivating example.}
We built a toy application, in which
every request executes an empty loop: \code{for(int i=0; i<max, i++)\{\}} (max=25K).
We disabled compiler
optimization in case it eliminates the loop. 
Our initial purpose to build this application
is to create a simple baseline without
any internal variance, so that we can use it to measure
variance caused by external events like scheduling and interrupts.
Surprisingly, however,
it incurs a non-negligible variance: while 50\% of the requests have a latency of about 60us, the remaining
ones can reach 90us. We had a few guesses about the cause, but without a proper tool, we were
not able to confirm the cause. In the following two sections,
we will present how \sys records and analyzes events using this example.

	\section{Low-overhead recording}
\label{sec:kernel}

\sys needs to record event information at runtime.
We first explore whether we can utilize existing
tools.
Perhaps not surprisingly, we find
extensive recording will incur a large I/O and space overhead:
when using \emph{ftrace} to record the timing of all
interrupt and thread scheduling events for a Memcached server,
it can
generate about 104MB of trace per second: this is about 9TB of trace per day.
Such result is reasonable since
Memcached can process millions of requests per second.
We have tried other applications and other recording tools, and find that
it is common for them to generate tens of MBs of trace per second (see Section~\ref{sec:overhead}). 
Such overhead is unlikely to be acceptable in the long term.

Such result suggests that we can only afford to record a small subset
of events. We call this mechanism \emph{selective recording} in this paper.
This observation is consistent with prior reports. For example, Google reports
that it records one out of 1024 requests in Dapper and considers further lowering
the rate~\cite{Sigelman2010Dapper}.
Selective recording brings the question which events we should record:
the subset of events we record must allow meaningful analysis, and
must fit with selective recording at the application level.

\subsection{\sys's approaches}
\label{sec:approaches}
To minimize recording overhead without disrupting the following analysis,
\sys follows two principles.

\vspace{.1in}
\noindent{\textbf{Design principle 1: record events that are affecting the selected requests.}
Because latency is defined on the granularity of application requests, eventually \sys's offline analysis
has to answer the question which events are contributing to the latency of different requests.
Therefore, if the application selectively records a subset of requests, \sys should record
kernel and hardware events that are affecting these requests; otherwise, \sys may record events that are
useless in the following offline analysis.

Existing kernel tools can record events happened on the whole machine, a particular
CPU core, a particular process, or a particular thread~\cite{perf,ftrace,kprobe}. We find recording
events at thread granularity fits well with our first principle, due to the following reason: as mentioned
in Section~\ref{sec:related}, application-level tools usually model a request as multiple
\utasks; when they select
a request, they will record the \utasks created by
the request. Since such \utasks are usually defined as a block of code executed on a thread,
recording kernel and hardware events at thread granularity is a natural fit.
To be more precise, \sys should provide APIs to start and stop recording kernel and hardware
events happening on any particular thread; the application-level
tools can then use such APIs to record events affecting a selected \utask.

\vspace{.1in}
\noindent{\textbf{Design principle 2: record coarse-grained information first and record
other information only when necessary.}
Existing tools can record the timing of each event, the call stack of each event,
or the cumulative values of the same type of events~\cite{perf,ftrace}. Recording cumulative
values has the least I/O overhead per event, considering an \utask may be affected by
multiple events of the same type.
Furthermore, it has a nice property that
its required buffer size per thread is bounded by the number of types of events to record.

Compared to recording timing or call stacks,
recording cumulative values will lose some information, but we find such loss is either fine or can be compensated by additional
mechanisms.
First, recording cumulative values will lose the timing
of each individual event. This is fine since events like interrupts are not caused by the \utask to monitor,
so the timing of these events is not much helpful anyway; for those kernel
events caused by \utasks, such as waiting for lock or I/O, they can be
captured by application-level recording.

Second, compared to recording call stacks, recording cumulative values
does not show the code related to the specific event.
This is fine in many cases, since many events like interrupts and thread scheduling
are self explanatory. For those events which need to be connected
to code (e.g., cache miss), we can sample call stacks after we locate the
problematic events. 
Compared to recording call stacks all the time, \sys's
approach, which is to record the cumulative values first and only sample the
call stacks of problematic events, can significantly reduce the overhead.

\vspace{.1in}
\noindent{\textbf{Challenges.}} While existing tools have provided some of the features
mentioned above, we find two important functions are missing.
First, when recording cumulative values, we find both the cumulative count
and the cumulative length (i.e., wall-clock time) are useful: longer latency could be caused by
a longer event and/or more events. However, existing tools only record cumulative
counts. Recording cumulative length accurately is actually more challenging than
we thought: in the OS kernel, an event can be preempted by others, so it is challenging
to match the end of an event to the beginning of the event, and it is inappropriate
to just use the end time minus start time to compute the length of the event.

Second, when recording the count of events happened on a given thread, existing tools do
not separate the change of the count caused by interrupts from that caused by thread execution.
This is not ideal since we may double count the effect of interrupts. For example,
if we observe a long-latency request has both high interrupt count and high
memory access count, are those additional memory accesses caused by the request
itself or by the additional interrupts?

Next we present the design and implementation of \sys's recording module, which follows the design
principles mentioned above and adds the missing pieces.

\subsection{\sys API}
\label{sec:utask}

Following the principle to record events
at thread granularity,  \sys provides two APIs functions,
\code{begin\_AppTask(taskID)} and \code{end\_AppTask(taskID)},
to allow a developer or an application level tool to define
\utasks:
any continuous piece
of code within one thread can be defined as an \utask.
\sys uses the \code{taskID} argument to differentiate
different \utasks.

\begin{figure}[tb]
\centering
 \pseudocodeinput{example_API}
  \vspace{-.1in}
 \caption{Usage of \sys API.}
 \label{fig:example-api}
 \vspace{-.2in}
 \end{figure}
 
Figure~\ref{fig:example-api} presents
how to integrate \sys with application level recording in
the example we discussed in Section~\ref{sec:overview}.
This example defines an empty loop as a request (lines 7).
Suppose the developer prints the timing
of the empty loop to measure its latency variance (lines 4 and 10). 
To limit overhead, the developer decides to record a subset of them (lines 2, 3, and 9).
To apply \sys, the developer
simply needs to append the \code{begin\_AppTask} and \code{end\_AppTask}
calls to the existing annotation (lines 5 and 11).

Once again, as shown in this example, \sys does not answer the
questions about where to add annotation and which requests to select, which are
well studied in existing application level tools. Instead, \sys enhances these
tools to record and analyze kernel and hardware events.

\subsection{Recording cumulative values}
\label{sec:utaskimpl}

This section presents how \sys records the cumulative
values of events at run time with the example in Figure~\ref{fig:example}:
it executes three loops as shown in Figure~\ref{fig:example-api};
loop1 and loop3 are executed on thread T1; loop2 is executed on
thread T2; loop2 and loop3 are selected for recording but loop1 is not.

As discussed in Section~\ref{sec:approaches}, 
\sys inherits designs from existing tools and
adds the missing pieces. For completeness, we present the whole
design of \sys but we emphasize the parts that are different from
existing tools.

\vspace{.1in}
\noindent{\textbf{Maintaining context metadata.} Similar
as existing tools like \emph{perf}~\cite{perf}, \sys maintains a context metadata for each
thread and for each CPU core. The thread context metadata (TCM) contains
a flag indicating whether recording is enabled for this thread
and a vector to record event values; the core context metadata (CCM)
contains the ID of the thread running on the corresponding
core and inherits the recording flag from the TCM of the thread. \code{begin\_AppTask} and \code{end\_AppTask}
will initialize and clean TCM and CCM;
when a thread is scheduled to run on a particular core, \sys will update the core's CCM
based on the thread's TCM.

For example, in Figure~\ref{fig:example}, since loop2 and loop3 are selected for
recording, their \code{begin\_AppTask} and \code{end\_AppTask} will be executed:
at $t_3$, the \code{begin\_AppTask} of loop2 will set \code{T2.TCM.recording} and
\code{Core2.CCM.recording} to be true and at $t_4$, \code{end\_AppTask} of
loop2 will set both flags to false; loop3's  \code{begin\_AppTask} and \code{end\_AppTask}
will perform the similar execution; in addition, when T1 is scheduled out at $t_7$,
\sys will set \code{Core1.CCM.recording} to false since T1 is no longer running on Core1;
when T1 is scheduled to run on Core2 at $t_8$, \sys will set \code{Core2.CCM.recording} to
be equal to \code{T1.TCM.recording} since now T1 is running on Core2.

When a kernel event occurs, \sys
checks its corresponding TCM or CCM to find whether recording is enabled: if yes, \sys will process the event as follows:

\begin{figure}[t]
\centering
\includegraphics[width=0.43\textwidth]{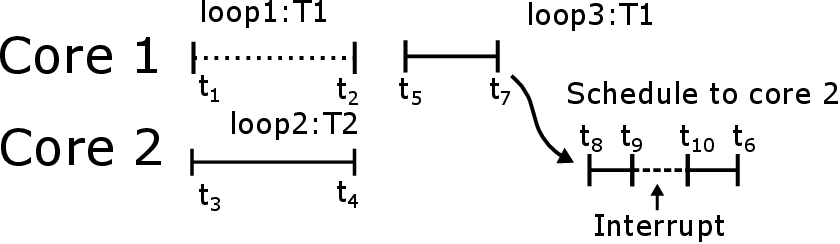}
\vspace{-.1in}
\caption{An example of executing multiple loops (loop1 and loop3 run on thread T1; loop2 runs on thread T2; loop2 and loop3 are selected for recording).}
\vspace{-.25in}
\label{fig:example}
\end{figure}

\vspace{.1in}
\noindent{\textbf{Recording the lengths of scheduling events.}
If thread $A$ was scheduled to wait (either RUNNABLE or BLOCKED) at time $t_1$ and scheduled to run
at time $t_2$ while an \utask $T$ was running on thread $A$, \sys will add $t_2 - t_1$ to the
cumulative length of scheduling events of $T$.

\sys's implementation is similar to existing tools: \sys maintains a \code{last\_wait\_time} variable in the TCM of each thread;
\sys updates this variable when the thread
is scheduled to wait (\code{last\_wait\_time = time()}); and when the thread is scheduled to run, \sys computes the length of
scheduling (\code{length = time()-last\_wait\_time}) and adds the length to the corresponding counter in the TCM.

For example, in Figure~\ref{fig:example}, at $t_7$, since T1 is scheduled out, \sys will set
\code{T1.TCM.last\_wait\_time} to $t_7$; then at $t_8$, since T1 is scheduled to run, \sys will
add $t_8$-\code{T1.TCM.last\_}
\code{wait\_time} = $t_8-t_7$ to the counter in \code{T1.TCM}.

\vspace{.1in}
\noindent{\textbf{Recording the lengths of interrupts and faults.} 
This is more challenging than recording the lengths of scheduling because interrupt or fault can be preempted
by other interrupts, possibly multiple times, and such preemption can
be nested. This creates challenges for matching the end and the start of an
event and computing its precise length.
As far as we know, no existing tools record the precise lengths of interrupts
and faults at runtime. 

\sys's solution is based on the following property:
if an event $e_1$ is preempted by $e_2$, then $e_2$ must finish before $e_1$.
This is because OS maintains interrupts contexts using stacks and
will have to finish later interrupts first.
This property simplifies our problem in two ways: first, we can match the end of an event to
the start of the event by simulating the stack maintained by the OS, like matching right parenthesis to left
parenthesis; 
second,
if an event $e_1$ is preempted by $e_2$ from $t_1$ to $t_2$, we can simply subtract
$t_2 - t_1$ from $e_1$'s length without considering whether $e_2$ was preempted
by others, because the property mentioned above guarantees that any events preempting
$e_2$ must happen between $t_1$ and $t_2$.

\begin{figure}[t]

\centering
\pseudocodeinput[]{pseudo_interrupts.tex}
\vspace{-.1in}
\caption{Recording the lengths of interrupts and faults.}

\label{fig:interrupts}
\vspace{-.15in}
\end{figure}

Figure~\ref{fig:interrupts} shows \sys's solution. \sys maintains a per-core shadow stack to simulate
how the OS keeps track of preemption relationships (lines 1-5).
Each level in the stack records the start time of the corresponding interrupt/fault
(line 2) and the aggregate length of it being preempted (line 3).
When an interrupt/fault starts, \sys pushes a new level to the stack (lines 8-11);
when an interrupt/fault ends, \sys can know it should be matched to the top of the
stack, based on the property mentioned above (line 15). \sys computes both the length including
preemption (line 16) and the length excluding preemption (line 17):
the latter is the precise length of the interrupt/fault and thus will be added to the
cumulative counter in the TCM (line 20); the former will be used to update the
information of the interrupt/fault being preempted by the current one (lines 18-19).

For example, in Figure~\ref{fig:example}, when interrupt happens at $t_9$,
\sys will push one level into Core2's shadow stack; when interrupt ends
at $t_{10}$, \sys will pop the top level from Core2's shadow stack and compute
its total length ($t_{10}-t_9$); since this interrupt is not preempted by others, its preempted\_length
is 0, so $t_{10}-t_9$ is its actual length and \sys will add this value to \code{T1.TCM}.

\vspace{.1in}
\noindent{\textbf{Recording the values of CPU performance counters.} 
At a high level, \sys's solution is not different from existing tools:
\sys maintains
a \code{start\_value} variable in the TCM of each thread to record
the counter value when the thread is scheduled to run (\code{start\_value = read\_counter()}); when the
thread is scheduled out, \sys computes the change of the value (\code{change = read\_counter()-start\_value})
and adds the change to the counter in the TCM.

However, \sys differs from existing tools in that it
excludes the change of counters caused by executing
an interrupt. To achieve that, we re-define the concept
of ``running'' in the following way:
an \utask starts to run
when the \utask starts (i.e., \code{begin\_AppTask}),
when the \utask gets scheduled in, or when an interrupt/fault
preempting the \utask finishes; the \utask stops to run
when the \utask ends (i.e., \code{end\_AppTask}),
when the \utask gets scheduled out, or when an interrupt/fault
preempts the \utask. This definition essentially treats
interrupt preemption in the same way as the thread
getting scheduled out. Based on this definition,
we can re-use the solution mentioned
above by adding the corresponding logic
when interrupt starts or ends.

For example, in Figure~\ref{fig:example}, for loop3,
\sys will compute the change of counters in the periods
of $t_5$ to $t_7$, $t_8$ to $t_9$, and $t_{10}$ to $t_6$,
and add the value of each period to \code{T1.TCM}.

\vspace{.1in}
\noindent{\textbf{Writing values to persistent storage.}
\code{end\_AppTask} writes the latency of the
\utask and its counter values stored in the TCM to a trace file,
which will be used for offline analysis.

\vspace{.1in}
\noindent{\textbf{Implementation and optimizations.} 
We implement \sys's recording module with 1426 lines of code
and adds 563 lines of code in Linux kernel. We tried to avoid kernel
modification by utilizing kernel instrumentation tools like \emph{kprobe}
and \emph{eBPF}, but we find
they rely on predefined trace
points in the kernel and a few functions we need
have not been defined as trace points yet (e.g. \textit{do\_page\_fault()}). 
Therefore, we had to modify kernel source code.

\sys incorporates a number of optimizations to minimize overhead. First,
since \code{begin\_AppTask} and \code{end\_AppTask} are executed in the user space,
they need systems calls to read CPU counters.
To minimize the number of system calls, \sys's kernel module exports an \emph{ioctl} system call to execute all kernel-related logic of
\code{begin\_AppTask} and \code{end\_AppTask}.

Second, to maximize lookup efficiency, \sys uses an array indexed by thread ID or core ID to store the TCMs and CCMs.
Since the maximal thread ID can be large, \sys divides the TCM array into blocks, each
for a range of 64K threads, and allocates a block when a thread in its range is actually
recorded. In practice, we observe the thread IDs of a process are usually close, and thus
they will require only one or a few blocks.
\sys currently includes 14 counters in total, including eight for kernel events and
six for CPU events, and thus each block requires about 7MB of memory.

Finally, \sys can only record a subset of events for each \utask,
because the CPU only allows to record a subset of counters (see Section~\ref{sec:related}).
Without a prior knowledge of which events are problematic, \sys has to
try different events for different \utasks.
Changing what to record incurs overhead to store the types of events to record.
Therefore, instead of making a selection for
each \utask, \sys
records the same types of events for \utasks in a short period of time, called an \emph{epoch};
\sys only changes what to record between epochs and thus amortizes the
overhead of switching across all \utasks in an epoch. In our experiments,
we set epoch length to be one second, which has negligible overhead.

\vspace{.1in}
\noindent{\textbf{Overhead of recording.}
\sys stores a fixed amount of metadata per thread and per
core.
The computation overhead for each event is \emph{O(1)} since no step needs
to scan multiple events. Furthermore, note that if an \utask is not selected for recording, then
its corresponding events can avoid most of the processing discussed above,
except that a scheduling event still needs to update the \code{recording} flag and thread ID in the affected TCM and CCM.

	\section{Identifying the causes of variance}
\label{sec:statistics}

\sys's online recorder will generate a trace file, which contains the latency of each selected \utask
and a set of cumulative values for different events associated with the corresponding \utasks.
However, because of selective recording, different \utasks may have the values for a different
set of events.
For example, for Figure~\ref{fig:example}, the recorded trace file may look like the following:

\vspace{.1in}
\noindent \textbf{trace file:}

\begin{small}
\noindent loop2 (100us): running=100us, sched=0us, interrupt=0us, TLB miss=3

\noindent loop3 (150us): running=100us, sched=20us, interrupt=30us, L1 miss=0

\vspace{.1in}
\end{small}
	
The goal of \sys's offline analyzer is to find which events are causing latency variance.
Note that there is a subtle difference between finding events whose \emph{variance} contributes
significantly to the variance of \utask latency---this is the goal of \sys---and finding events which contributes
significantly to long-latency \utasks. For example, in the trace shown above, the execution of
the loop (100us) is contributing a lot to the latency of loop3, but it is not the
reason why loop3 takes longer than loop2, because for both loop2 and loop3, their
execution of the loop takes 100us.

At the first glance, this looks like a classic statistical problem, but our investigation
of existing methods encounters challenges due to the specific properties and requirements of our problem.
To motivate \sys's solution, we first present the properties we find necessary and
how existing solutions violate these properties.

\begin{itemize}[leftmargin=*]

\vspace{-.05in}
\item Efficient on large data set: despite selective recording, recording for a long time will generate
a large amount of trace. 
As a result, we find any analysis that has $O(N^2)$ complexity (N is the number of requests) or more
is time consuming and not preferred. Therefore, we have to avoid complicated analysis.

\item Informative to the user: for the purpose of reducing latency variance, \sys should provide a quantitative estimate about the
benefit of optimizing each type of event. This goal is complicated by the fact that CPU events
are high correlated, which means we need to infer the causal relationship among different events.
Among the methods we investigated,
correlation analysis~\cite{benesty2009pearson,myers2004spearman}
can determine whether two variables are related, but they don't
estimate the actual impact of a variable: an event causing the latency
to increase by 1ms may have the same correlation value as one causing the latency to increase by 10ms.
To infer causal relationship, statistical methods usually rely on controlled experiments~\cite{correlationcausation,causalanalysis},
which change the value of one event to see how it affects the other. In \sys, however, this is
usually infeasible: we cannot arbitrarily insert an event like a cache miss into an \utask.

\item Robust on different data set and against missing data: our analysis should minimize any mathematical assumptions about
the data set because those assumptions may not hold. Furthermore,
since each request only has values for a subset of events, our analysis should be able
to work with such ``missing data''. Among the methods we investigated, regression
methods usually rely on mathematical assumptions like linear relationship; some
also rely on the assumption that data is not correlated; methods like Principal Component Analysis (PCA)~\cite{wold1987principal}
can decouple the correlation among different events, but they cannot work with
missing data.

\end{itemize}

Such exploration motivates us to develop new efficient, informative, and robust methods to achieve our goal.

\subsection{Computing the impact of events}
\label{sec:highimpact}

To compute the impact of an event $E$ on the latency of \utask $T$, our key idea is to simulate
the effect of optimizing $E$ by answering the following question:
if we could reduce the variance of $E$, how much could it reduce the
variance of $T$'s latency? To answer this question,
we could compare the latency variance of $T$s with and without
high values of $E$.

Following this idea, \sys defines the \emph{impact value} of event $E$
on \utask $T$ as follows:

\begin{definition}
$impact(E)_T = \frac{\mathit{Var}(\vec{T_E})-\mathit{Var}(\vec{T_E} - \vec{T_{hE}})}{\mathit{Var}(\vec{T_E})}$
\end{definition} 

In this definition, $\mathit{Var}$ is a function to compute
the variance of a set of values (more precisely a multiset since
there might be duplicate values); $\vec{T_E}$ is a set containing the latencies of all
\utasks which have recorded $E$; and $\vec{T_{hE}}$ is a set containing the latencies of \utasks with high values
of $E$. Intuitively, this definition computes, if we could eliminate
$E$ with high values, how much could it reduce the variance of $T$'s latency.

This definition has three benefits: first, it
estimates the actual benefit of optimizing a certain event. 
Second, it does not rely on any mathematical assumptions of
the data. Third,
it directly tells us how to compute its value, as long as we can define $\mathit{Var}$ and determine the threshold
of high $E$ value.

\vspace{.1in}
\noindent{\textbf{Measure the variance.}
To quantitatively measure the variance of \utask latencies (i.e., the $\mathit{Var}$ function), prior work
either computes the Coefficient of Variation (CoV) of their values~\cite{Maricq2018Variability}, or use the
tail value at a certain percentile~\cite{Decandia07Dynamo,Huang2017SAL,Berger2018RobinHood,Li2014Tales}. \sys uses the tail value since it is
commonly used in existing systems and CoV is not very sensitive to variance within a small percentile of values.
\sys introduces a parameter \emph{pTarget} to describe the target percentile. For example, \emph{pTarget}=0.99 means
using 99 percentile latency to measure the variance. In practice, we expect the developer
to set the value of \emph{pTarget}, either depending on the CDF of the latency or depending on the requirement of the system.

\vspace{.1in}
\noindent{\textbf{Determine the threshold of high event value.}
\sys introduces a parameter \emph{pThreshold}
to determine the threshold of high $E$ value.
For example, \emph{pThreshold}=0.8 means values over the p80
value of $E$ are considered ``high''.
To find this threshold, we observe that the CDF of event values sometimes
look like ``steps'' (e.g., Figure~\ref{fig:cpubench-cdf}), which indicates a significant change at
these steps. In this case, choosing the percentile at these steps is an obvious choice.

Following this idea, \sys introduces a method to automatically find such step points. It divides the CDF of
the target event into N ranges (N=1000 in our experiments) and performs linear
regression for each range. Then it tries to merge adjacent ranges by performing linear
regression on the combination of them: if the R-squared value, which is used to measure how successful
the linear regression is, is large than 0.95 for both ranges under the new regression model,
\sys considers the merge as successful. By using this approach, \sys can
use a number of lines to fit the CDF of the target event, and the connection points of
adjacent lines are our targets: \sys will choose the one whose percentile
is smaller than \emph{pTarget} and is the closest to \emph{pTarget} as \emph{pThreshold}.

Given linear regression models on two ranges,
performing linear regression on the combination of them can be done
in $O(1)$ time, by re-using the intermediate results from the previous two
regression models~\cite{apache_2016}. This property significantly reduces the computation
overhead of our method.

\vspace{-.05in}
\subsection{Inferring causal relationship}
\label{sec:rules}

Many types of events \sys records are potentially correlated.
For example, if an \utask executes more instructions, it is likely
to cause more cache hits/misses.
High correlation among events
brings a challenge to our analysis: an event with a high impact
value does not mean it is worth optimizing because its impact could be caused
by another correlated event.
To address this challenge, \sys introduces two domain-specific rules to infer the causal
relationship between events and adjust their impact values.

The first rule infers the causal relationship based on the types of events.
We classify CPU events into five groups: CYCLE, INST, CACHE, UOP, and OTHER.
CYCLE includes events measured in the number of cycles; INST includes events measured in
the number of instructions; CACHE includes events to measure cache and TLB hits and misses; UOP
includes micro-architecture events, like decoding an instruction into micro-ops; OTHER only contains one event, which is the number of hardware interrupts. 
\ifdefined\TR
The event classification is shown in Table~\ref{table:cpu-events}.
\begin{table*}[htbp]
	\centering
	\small
	\begin{tabular}{@{}llll@{}}
		\toprule
		Event name & Group & Event name & Group \\ \midrule
		ITLB.ITLB\_FLUSH  & CACHE & FP\_ASSIST.ANY  & CYCLE \\
		TLB\_FLUSH.DTLB\_THREAD  & CACHE & CPU\_CLK\_UNHALTED.THREAD\_P  & CYCLE \\ 
		MEM\_INST\_RETIRED.SPLIT\_LOADS  & CACHE & CPU\_CLK\_UNHALTED.THREAD\_P\_ANY & CYCLE \\ 
		MEM\_INST\_RETIRED.SPLIT\_STORES & CACHE & CPU\_CLK\_THREAD\_UNHALTED.REF\_XCLK  & CYCLE \\ 
		MEM\_LOAD\_RETIRED.L1\_HIT  & CACHE & CPU\_CLK\_THREAD\_UNHALTED.REF\_XCLK\_ANY & CYCLE \\ 
		MEM\_LOAD\_RETIRED.L2\_HIT  & CACHE & CPU\_CLK\_THREAD\_UNHALTED.ONE\_THREAD\_ACTIVE  & CYCLE \\ 
		MEM\_LOAD\_RETIRED.L3\_HIT  & CACHE & L1D\_PEND\_MISS.PENDING\_CYCLES & CYCLE \\ 
		MEM\_LOAD\_RETIRED.L1\_MISS & CACHE & L1D\_PEND\_MISS.PENDING\_CYCLES\_ANY  & CYCLE \\ 
		MEM\_LOAD\_RETIRED.L2\_MISS & CACHE & DTLB\_STORE\_MISSES.WALK\_PENDING & CYCLE \\ 
		MEM\_LOAD\_RETIRED.L3\_MISS & CACHE & DTLB\_STORE\_MISSES.WALK\_ACTIVE  & CYCLE \\ 
		L2\_TRANS.L2\_WB  & CACHE & EPT.WALK\_PENDING & CYCLE \\ 
		L2\_LINES\_IN.ALL & CACHE & DTLB\_LOAD\_MISSES.WALK\_PENDING  & CYCLE \\ 
		LONGEST\_LAT\_CACHE.REFERENCE & CACHE & DTLB\_LOAD\_MISSES.WALK\_ACTIVE & CYCLE \\ 
		LONGEST\_LAT\_CACHE.MISS  & CACHE & LOCK\_CYCLES.CACHE\_LOCK\_DURATION  & CYCLE \\ 
		L1D\_PEND\_MISS.PENDING & CACHE & ICACHE\_64B.IFDATA\_STALL & CYCLE \\ 
		DTLB\_LOAD\_MISSES.WALK\_COMPLETED  & CACHE & ITLB\_MISSES.WALK\_PENDING  & CYCLE \\ 
		DTLB\_STORE\_MISSES.MISS\_CAUSES\_A\_WALK & CACHE & ILD\_STALL.LCP  & CYCLE \\ 
		DTLB\_STORE\_MISSES.WALK\_COMPLETED & CACHE & INST\_RETIRED.ANY\_P  & INST  \\ 
		DTLB\_STORE\_MISSES.STLB\_HIT & CACHE & BR\_INST\_RETIRED.ALL\_BRANCHES & INST  \\ 
		L1D.REPLACEMENT & CACHE & BR\_INST\_RETIRED.CONDITIONAL & INST  \\ 
		DTLB\_LOAD\_MISSES.STLB\_HIT  & CACHE & BR\_INST\_RETIRED.NEAR\_CALL  & INST  \\ 
		ICACHE\_64B.IFTAG\_HIT  & CACHE & BR\_INST\_RETIRED.ALL\_BRANCHES & INST  \\ 
		ICACHE\_64B.IFTAG\_MISS & CACHE & BR\_INST\_RETIRED.NEAR\_RETURN  & INST  \\ 
		ITLB\_MISSES.MISS\_CAUSES\_A\_WALK  & CACHE & BR\_INST\_RETIRED.NOT\_TAKEN  & INST  \\ 
		ITLB\_MISSES.WALK\_COMPLETED  & CACHE & BR\_INST\_RETIRED.NEAR\_TAKEN & INST  \\ 
		ITLB\_MISSES.STLB\_HIT  & CACHE & BR\_INST\_RETIRED.FAR\_BRANCH & INST  \\ 
		MEM\_INST\_RETIRED.STLB\_MISS\_LOADS  & CACHE & BR\_MISP\_RETIRED.ALL\_BRANCHES & INST  \\ 
		MEM\_INST\_RETIRED.STLB\_MISS\_STORES  & CACHE & BR\_MISP\_RETIRED.CONDITIONAL & INST  \\ 
		L2\_RQSTS.CODE\_RD\_MISS  & CACHE & BR\_MISP\_RETIRED.ALL\_BRANCHES & INST  \\ 
		RESOURCE\_STALLS.ANY  & CYCLE & BR\_MISP\_RETIRED.NEAR\_TAKEN & INST  \\ 
		RESOURCE\_STALLS.SB & CYCLE & FP\_ARITH\_INST\_RETIRED.SCALAR\_DOUBLE & INST  \\ 
		CYCLE\_ACTIVITY.CYCLES\_L2\_MISS  & CYCLE & FP\_ARITH\_INST\_RETIRED.SCALAR\_SINGLE & INST  \\ 
		CYCLE\_ACTIVITY.CYCLES\_L3\_MISS  & CYCLE & FP\_ARITH\_INST\_RETIRED.128B\_PACKED\_DOUBLE & INST  \\ 
		CYCLE\_ACTIVITY.STALLS\_TOTAL & CYCLE & FP\_ARITH\_INST\_RETIRED.128B\_PACKED\_SINGLE & INST  \\ 
		CYCLE\_ACTIVITY.STALLS\_L2\_MISS  & CYCLE & FP\_ARITH\_INST\_RETIRED.256B\_PACKED\_DOUBLE & INST  \\ 
		CYCLE\_ACTIVITY.STALLS\_L3\_MISS  & CYCLE & FP\_ARITH\_INST\_RETIRED.256B\_PACKED\_SINGLE & INST  \\ 
		CYCLE\_ACTIVITY.CYCLES\_L1D\_MISS & CYCLE & MEM\_INST\_RETIRED.LOCK\_LOADS & INST  \\ 
		CYCLE\_ACTIVITY.STALLS\_L1D\_MISS & CYCLE & MEM\_INST\_RETIRED.ALL\_LOADS  & INST  \\ 
		CYCLE\_ACTIVITY.CYCLES\_MEM\_ANY  & CYCLE & MEM\_INST\_RETIRED.ALL\_STORES & INST  \\ 
		CYCLE\_ACTIVITY.STALLS\_MEM\_ANY  & CYCLE & HW\_INTERRUPTS.RECEIVED  & OTHER  \\ \bottomrule
	\end{tabular}
	\caption{Classification of CPU events.}
	\label{table:cpu-events}
\end{table*}
\else
We document the classification in Table 1 of an extended version~\cite{VarMRI2021TR}.
\fi

We ignore UOP events, because
software developers may not have a good understanding of them. We ignore OTHER as well because
\sys already records the length and count of hardware interrupt events.
As a result, \sys records 79 performance counters of groups CYCLE, INST, and CACHE.

\vspace{.1in}
\noindent{\textbf{Rule 1.} INST > CACHE > CYCLE, which means that if an event in group $G_1$ is correlated with an event in group $G_2$
($G_1$ and $G_2$ could be INST, CACHE, or CYCLE), 
and $G_1$ > $G_2$, then the former event is likely to be the cause.

The reason is as follows: taking INST > CYCLE as an example, if an \utask occasionally executes extra instructions, then it
may lead to extra cycles. On the contrary, if certain instructions of the \utask occasionally
require extra cycles, it usually will not affect the \utask's number of instructions,
because the logic of an \utask usually does not depend on how many cycles it takes.
One can make similar arguments for INST > CACHE and CACHE > CYCLE.

Based on Rule 1, \sys adjusts
the impact values of different events in the following way:
if $E_1$ may cause $E_2$ as defined in Rule 1, then $impact(E_2)_T -= impact(E_1)_T * Corr(E_1, E_2)_T$.
Intuitively, this means we should deduct the impact of $E_1$ from the impact of $E_2$, based on
how much they are actually correlated.
If multiple events may cause $E_2$,
then $impact(E_2)_T -= Max(impact(E_i)_T * Corr(E_i, E_2)_T)$ ($E_i$ may cause $E_2$):
here we deduct the maximal one as an approximate solution, since we don't know whether $E_i$s are correlated, 
and thus we cannot deduct
the impact of each $E_i$.

To compute the correlation of two types of events $E_1$ and $E_2$, \sys first finds the \utasks which have recorded
both types together. Among them, \sys finds the groups of \utasks with high values of $E_1$ and $E_2$, denoted as $T_{E1}$
and $T_{E2}$. \sys finally computes the 
Jaccard Distance~\cite{Paul1901Jaccard} of these two groups $\frac{| T_{E1} \cap T_{E2} |}{| T_{E1} \cup T_{E2} |}$,
and use the result as the correlation value.
Its basic idea is that, if two types of
events are highly correlated, then the \utasks with high values of each type should overlap significantly.

The second rule tries to infer the causal relationship between
events of the same type.

\begin{definition}
 An event $E_a$ is the child of another event $E_b$ if the value of $a$ is included in the value of $b$.
\end{definition}

Such relationship is common in CPU performance counters. For example, the
counter to record total stalled cycles is a child of the counter to record total cycles.
\ifdefined\TR
We summarize the event pairs in Table~\ref{table:parent-child-events}.
\begin{table*}[htbp]
	\small
	\begin{tabular}{@{}ll@{}}
	\toprule
	\centering
		Child event & Parent event \\ \midrule
		CYCLE\_ACTIVITY.STALLS\_L2\_MISS & CYCLE\_ACTIVITY.STALLS\_TOTAL \\ 
		CYCLE\_ACTIVITY.STALLS\_L3\_MISS & CYCLE\_ACTIVITY.STALLS\_TOTAL \\ 
		CYCLE\_ACTIVITY.STALLS\_L1D\_MISS & CYCLE\_ACTIVITY.STALLS\_TOTAL \\
		CYCLE\_ACTIVITY.STALLS\_MEM\_ANY	&	CYCLE\_ACTIVITY.STALLS\_TOTAL	\\
		BR\_INST\_RETIRED.ALL\_BRANCHES & INST\_RETIRED.ANY\_P \\
		BR\_INST\_RETIRED.NOT\_TAKEN	&	BR\_INST\_RETIRED.ALL\_BRANCHES	\\ 
		BR\_INST\_RETIRED.FAR\_BRANCH	&	BR\_INST\_RETIRED.ALL\_BRANCHES	\\ 
		BR\_MISP\_RETIRED.ALL\_BRANCHES	&	BR\_INST\_RETIRED.ALL\_BRANCHES	\\ 
		BR\_MISP\_RETIRED.CONDITIONAL	&	BR\_INST\_RETIRED.ALL\_BRANCHES	\\ 
		BR\_MISP\_RETIRED.ALL\_BRANCHES	&	BR\_INST\_RETIRED.ALL\_BRANCHES	\\ 
		BR\_MISP\_RETIRED.NEAR\_TAKEN	&	BR\_INST\_RETIRED.ALL\_BRANCHES	\\ 
		MEM\_INST\_RETIRED.LOCK\_LOADS &	MEM\_INST\_RETIRED.ALL\_LOADS	\\ 
		CPU\_CLK\_UNHALTED.THREAD\_P	&	CPU\_CLK\_UNHALTED.THREAD\_P\_ANY \\
		CPU\_CLK\_THREAD\_UNHALTED.REF\_XCLK & CPU\_CLK\_THREAD\_UNHALTED.REF\_XCLK\_ANY \\
		CPU\_CLK\_THREAD\_UNHALTED.ONE\_THREAD\_ACTIVE & CPU\_CLK\_THREAD\_UNHALTED.REF\_XCLK\_ANY \\
		L1D\_PEND\_MISS.PENDING\_CYCLES & L1D\_PEND\_MISS.PENDING\_CYCLES\_ANY \\ \bottomrule
	\end{tabular}
	\caption{Summaries of parent-child relationship. Note that two obvious groups---all INST events are children
		of total instructions and all CYCLE events are children of total cycles---are omitted here to save space.}
	\label{table:parent-child-events}
\end{table*}
\else
We record all the pairs we found
in Table 2 of the extended version~\cite{VarMRI2021TR}.
\fi

\vspace{.1in}
\noindent{\textbf{Rule 2.} If $E_a$ is a child of $E_b$, and if $E_a$'s values are in direct proportion to $E_b$'s values (i.e., $E_b = constant * E_a$),
then $E_a$ is unlikely to be cause: it is likely
some other reason causes $E_b$ to increase, which causes $E_a$ to increase.

\ifx\TR\undefined
\vspace{.05in}
\fi
This rule is based on our empirical observation that if an event $E_b$ causes
its children's values to increase, then usually all $E_b$'s children's values will increase, often in a similar rate,
but if an event $E_a$ causes its parent's value to increase, usually there is no such effect to $E_a$'s siblings.

To apply this rule, 
\sys first finds \utasks which record a pair of such events together, and 
then applies linear regression on their values ($E_b=\alpha * E_a $): if the R-squared
value is higher than 0.99, \sys considers regression as successful and  removes $E_a$
from the report.

\vspace{.1in}
\noindent{\textbf{Output.}
Finally \sys will output all events ranked by their adjusted impact values.
If two events are correlated but neither rule is applicable, \sys will simply report the impact
values of these events and their correlation values.

\vspace{.1in}
\noindent{\textbf{Overhead of offline analysis.} The most time-consuming part of
\sys's offline analysis is to sort requests based on their latencies, whose complexity is $O(NlogN)$
($N$ is the number of requests). Other steps like line-based curve fitting and inferring
causal relationship have $O(N)$ complexity.

	\vspace{-.05in}
\subsection{Analysis of recording length}
\label{sec:combine}

Although selective recording is necessary to limit overhead,
we may miss rare events or need to record for longer
time to capture rare events. This section analyzes
such effect.


Suppose a problematic event
causes $\mathit{pTail}$ (in percentile) of the requests to have longer latencies; \sys records $\mathit{pReq}$ of the requests, and for these requests,
records $\mathit{pEvent}$ of the events. To capture the problematic event once, the application needs to process a total of $\frac{1}{\mathit{pTail} \times \mathit{pReq} \times \mathit{pEvent}}$
requests. In practice, we find setting $\mathit{pReq}$ to 1\% usually
will not incur a high overhead; $\mathit{pEvent}$ is around 5\% since CPU has 79
events and \sys can trace four of them at a time (number of kernel events is smaller); users with strict requirement about tail behavior may
set $\mathit{pTail}$ to 0.1\%~\cite{Decandia07Dynamo}; those add up to requiring
the application to run about two million requests in total to capture the problematic event once.
If we require the problematic event to occur, say, 100 times to reach a conclusion,
we need a total of 200 million requests.
For high-throughput applications like Memcached, which can process several
millions of requests per second, this does not require much time; for applications with lower throughput,
increasing $\mathit{pReq}$ is usually fine in terms of overhead but can decrease the number of required requests.

However, \sys exacerbates the problem because its rules to infer causal relationship require to analyze
many pairs of events, which requires these pairs to be recorded together. If we want to ensure \sys
can compute rules for any pair,  assuming a pair has chance $\mathit{pPair}$ to be recorded, this will change the previous
value to $\frac{1}{\mathit{pTail} \times \mathit{pReq} \times \mathit{pPair}}$. If \sys randomly selects events to record, $\mathit{pPair} = \frac{C^2_4}{C^2_{79}}$,
which is 26 times smaller than $\mathit{pEvent}$ (i.e., $\frac{4}{79}$). This means we need to record more requests.

Since each of our experiments lasts 16 hours, we find they can capture
sufficient number of events, despite such exacerbation. To shorten experiment time, we 
may differentially record more high-latency requests and fewer
median-latency requests, since high-latency requests are our targets~\cite{Sigelman2010Dapper}.

	\section{Case studies}
\label{sec:eval}

With the help of \sys, we have carried out a 3,000-hour study on two benchmarks
and four real-world applications on CloudLab~\cite{cloudlab}. It has generated 1.5TB of traces: 
without selective recording, it would have generated about 300 TBs of traces, which are
impossible to store and analyze for us.

Following a prior work~\cite{Maricq2018Variability}, we write a script to automatically 
allocate machines from CloudLab and run experiments on them. Each individual experiment
lasts 16 hours, which is the maximal allowed by CloudLab without extension (CloudLab only allows manual extension at this moment).
We randomly select an application and a workload to run in each experiment. We tune the
selection rate so that each experiment generates roughly 0.1MB of trace per second.

To verify the reports of \sys, we try to optimize the reported events and see whether the optimization
can reduce latency variance.
When simple optimization is not possible, we
search for works targeting the same problem.

\vspace{.1in}
\noindent{\textbf{Testbed.}  Our
machines are equipped with two 10-core Intel Xeon Silver 4114 processors, 192GB of memory, a 10Gb NIC, a 1TB hard disk, and Ubuntu 16.04 with Linux kernel 4.14.170.
We disable hyperthreading and turbo boost~\cite{Maricq2018Variability}.

\begin{table}[]
	\small
	\centering
	\begin{tabular}{@{}lccccc@{}}
		\toprule
		App & Segment info & Max & Median & Min &  Cov \\ \midrule
		Memcached-get & 238 $\times$ 2hrs & 86 & 45 & 39 & 12\% \\
		Memcached-set & 135 $\times$ 2hrs &76 & 64 & 61 & 4.4\% \\
		MySQL-YCSB & 104 $\times$ 3.5hrs &162 & 143 & 115 & 5.6\% \\
		ZK-get & 69 $\times$ 5hrs & 44 & 40 & 37 & 4.0\% \\
		ZK-set & 44 $\times$ 7hrs & 6.3e5 & 1.4e5 & 1.2e5 & 66\% \\
		HBase-set & 112 $\times$ 2hrs & 2090 & 2001 & 1937 & 1.7\% \\
		LoopBench & 189 $\times$ 2hrs & 86 & 84 & 82 & 1.0\% \\
		MemBench & 224 $\times$ 2hrs  & 12 & 11 & 11 & 2.0\% \\ \bottomrule
	\end{tabular}
	\caption{The statistics of p99 latencies across different segments. Max, median, and min
	are measured in us.}
	\vspace{-.4in}
	\label{table:tail-latency}
\end{table}

\vspace{.1in}
\noindent{\textbf{Analysis methodology.} Given a large volume of trace data, it's
time-consuming to analyze all of them. Instead, we use the
following strategy: we cut the
trace of each experiment into several segments, whose lengths are computed
based on the discussion in Section~\ref{sec:combine} (see actual values in Table~\ref{table:tail-latency}); we compute the
tail latencies of each segment to determine which segments are worth further analyzing.

As one can see in Table~\ref{table:tail-latency}, for some applications, their tail latencies can vary
significantly across different segments. For example, for Memcached-get, the max p99 latency
of different segments is 86us and the minimal p99 latency is 39us:
this is exactly why we need long-term recording to understand the causes of such variance.

To reduce analysis overhead, for segments with ``normal'' tail latencies,
we just analyze a few of them; then we mainly focus on those segments
with long tail latencies.
We try to answer the following two questions:

\begin{itemize}[leftmargin=*]

\item Within one segment, why is the tail latency longer than median latency?
To answer this question, we analyze
segments whose tail latencies are in the median range as shown in Table~\ref{table:tail-latency} (i.e., median-segment).

\item Across different segments, why do their tail latencies differ? To answer this question, we further analyze
segments with maximal tail latencies (i.e., max-segment).

\end{itemize}

\subsection{Loop bench}
First, we present the result on the toy benchmark as shown in Section~\ref{sec:overview}.
As a reminder, in this benchmark, each
 \utask executes an empty loop: for(int i=0; i<max, i++)\{\} (max=25K).
As shown in Figure~\ref{fig:cpubench-cdf}, while 50\% of the \utasks have a latency of about 60us, the remaining
ones can reach 90us. 

\sys (\emph{pTarget} = p85) reports that the CPU counter STALLS\_MEM\_ANY (execution stalls while memory subsystem has an outstanding load) is the top
cause. And the assembly code (Figure~\ref{fig:cpubench-original-assembly})
shows there is indeed a load instruction (i.e \textbf{mov}).
Its stall could be caused by waiting for the memory or cache
access to complete or by a data dependency in the pipeline (i.e., dependency on the \textbf{addl} instruction).
Since \sys reports almost no cache misses, cache access or pipeline stall are more likely.

To confirm the reason, we try to re-write the program and
find changing the code to for(long i=0; i<(int)max, i++)\{\} can remove most variance (Figure~\ref{fig:cpubench-cdf}).
As a side effect, the new code is faster than the original version.
The new assembly code (Figure~\ref{fig:cpubench-opt-assembly})
still has a load instruction, but it does not depend on the previous \textbf{addq} instruction, since their addresses are different.
This confirms that the pipeline stall is the cause of the variance, since the new version
still has cache accesses but has much less variance.
We further run \sys on the new version and find the STALLS\_MEM\_ANY counter
is not on top of the report any more.
While pipeline stall is known to be problematic for performance, it is unexpected to us that the latency of the stall can
have a non-negligible variance. 

Out of curiosity, we run the same benchmark on AMD EPYC 7452 and ARM APM X-GENE CPUs and find that 
the variance on these two types of CPUs is much smaller.
While this benchmark certainly does not represent any realistic workload, it shows an alarming
sign that some of today's hardware may have inherent performance variance despite running a deterministic
workload.

We don't observe significant cross-segment variance.

\begin{figure}[t]
	\centering
	\includegraphics[width=0.35\textwidth]{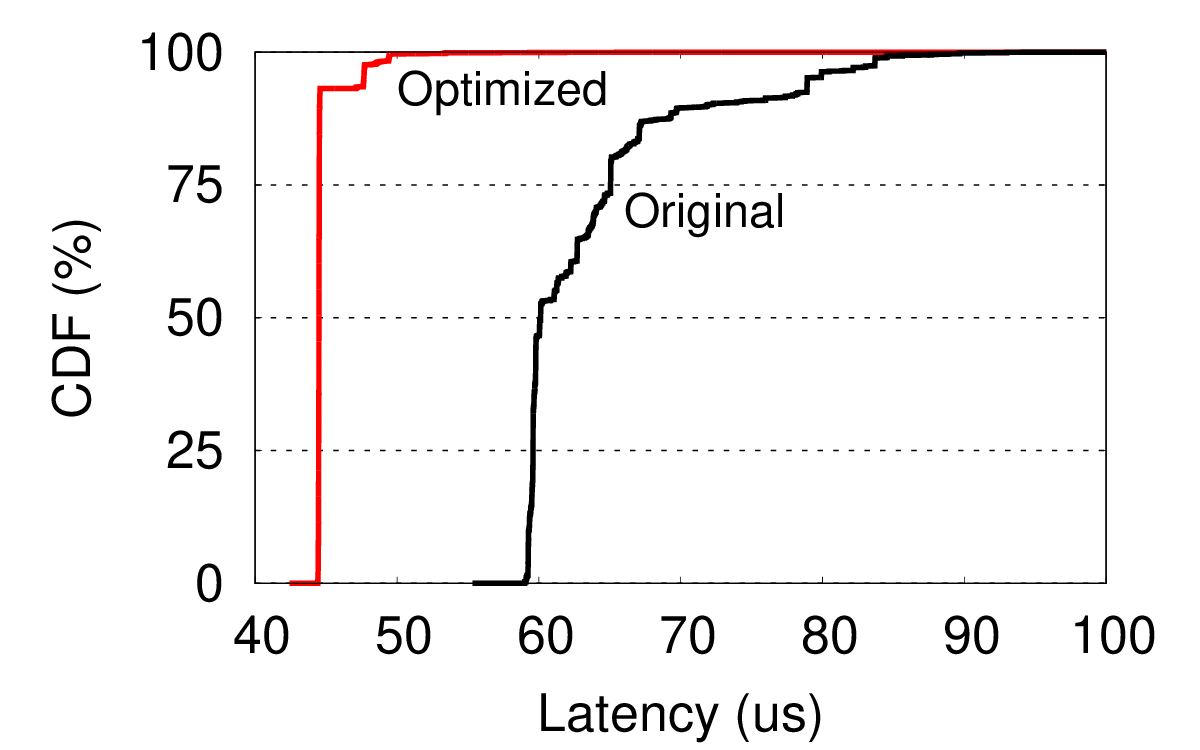}
	\vspace{-.1in}
	\caption{Loop bench experiments.}
	\label{fig:cpubench-cdf}
\end{figure}

\begin{figure}[t]
\centering
\begin{subfigure}[c]{0.45\columnwidth}
\begin{lstlisting}[frame=single, basicstyle=\footnotesize,linewidth=0.9\textwidth]
addl $0x1,-0x14(%rbp)
mov -0x14(%rbp),%eax
cmp -0xc(%rbp),%eax
jl to addl
\end{lstlisting}
\caption{Original version.}
\label{fig:cpubench-original-assembly}
\end{subfigure}
~
\begin{subfigure}[c]{0.45\columnwidth}
\begin{lstlisting}[frame=single, basicstyle=\footnotesize,,linewidth=0.9\textwidth]
addq $0x1,-0x8(%rbp)
mov -0xc(%rbp),%eax
cltq
cmp -0x8(%rbp),%rax
jg to addq
\end{lstlisting}
\caption{Optimized version.}
\label{fig:cpubench-opt-assembly}
\end{subfigure}
\vspace{-.1in}
\caption{Assembly code of Loop bench.}
\vspace{-.1in}
\end{figure}

\subsection{Memory bench} 
It creates enough pages to occupy almost
the full memory, so that these pages must be distributed to both sockets in our machine.
We define an \utask as accessing
all bytes in a random page, which is supposed to test memory latency and potential
NUMA effect. 

As shown in Figure~\ref{fig:membench-cdf},
about 25\% of the \utasks have a latency of about 8us and the remaining ones
can reach 10-12us.
We thought the variance is due to longer latency to access a remote socket,
but \sys (\emph{pTarget} = p25) reports a different
cause: page fault.

\sys uses \emph{perf} to sample the call stacks of page faults, and
finds they
are executing the \emph{migrate\_misplaced\_page} function, which is a technique to mitigate NUMA
effect. We disable this feature by setting \emph{/proc/sys/kernel/numa\_balancing} to 0,
and find the variance almost disappears: now almost all \utasks have a latency of 8us.
This means that the mechanism to mitigate NUMA effect can cause more
variance than the NUMA effect itself in certain cases. And as shown later, we find similar problems
in a number of real-world applications as well.
Of course, NUMA balancing is probably more beneficial when the workload has
better data locality.

We don't observe significant cross-segment variance.

\begin{figure}[t]
	\centering
	\includegraphics[width=0.35\textwidth]{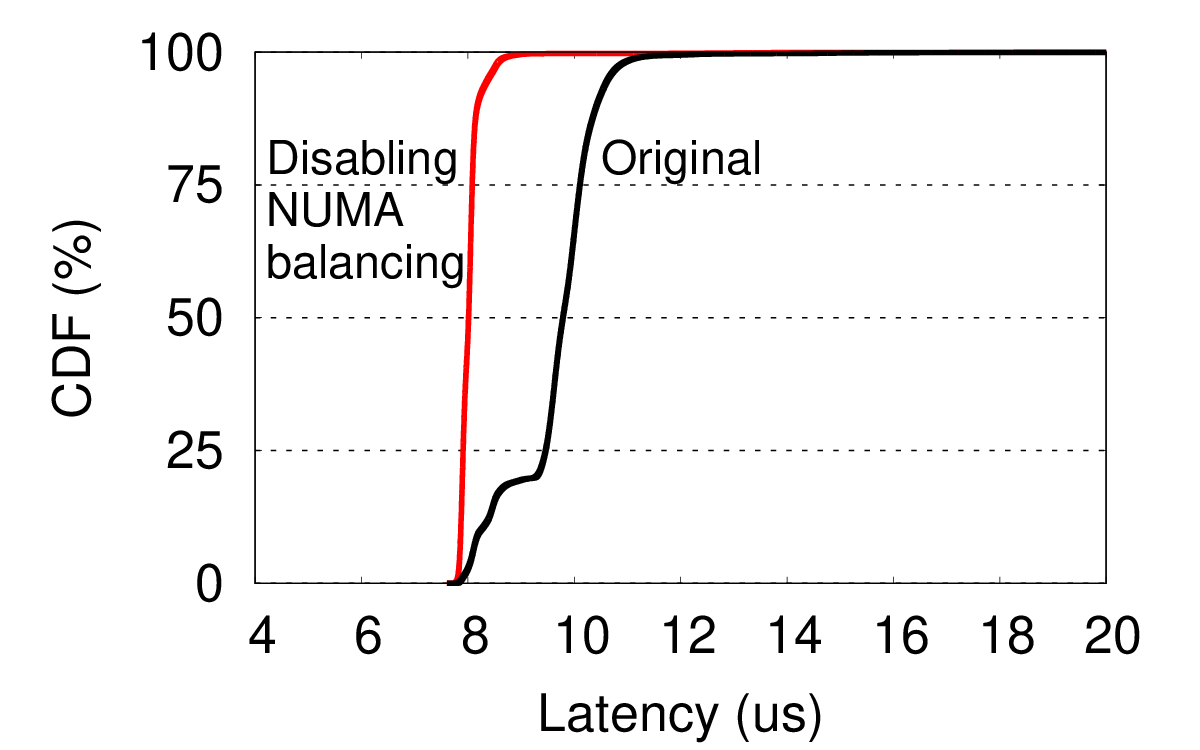}
	\caption{Memory bench experiments.}
	\vspace{-.1in}
	\label{fig:membench-cdf}
	\vspace{-.1in}
\end{figure}

\subsection{Memcached}
Memcached~\cite{memcached} is a in-memory
key-value store. It processes an entire request in one thread, so we define one \utask as the start
and the end of
the processing a request, i.e., we place \code{begin\_AppTask} after Memcached receives a request and
place \code{end\_AppTask} after Memcached sends the reply.

We test Memcached 1.4.36 with memaslap~\cite{memaslap}.
We configure memaslap to generate one million
key-value pairs with 16-byte key size and 16-byte value size and to generate a get-only workload and a set-only workload
both with random keys; we configure Memcached to use 20 threads; we test all
workloads under 80\% maximal throughput, which is a typical heavy load~\cite{Veera2016Kraken,Li2014Tales,Ousterhout2013Sparrow}.
We tried a mixed get-set workload as well but we do not find
anything new compared to get-only and set-only workloads, so we do not report it here.
We investigate the p99 latency (i.e., \emph{pTarget}=p99).

For the get-only workload, \sys reports the top cause of latency variance in the median-segment is NIC interrupts, i.e.,
an \utask may be preempted by interrupt handling to process an incoming packet;
the second cause is kernel instructions and \emph{perf} call stacks point
to the \emph{sendmsg} system call. Both are related to the network stack
and have been discussed in prior works, which eliminate such
overhead with kernel bypassing~\cite{Kalia2014RDMAKV,Lim2014MICA,Belady2014IX}.

We observe a significant cross-segment variance for this test  (Figure~\ref{fig:memcached-variance}): while about 220 segments have p99 latencies of 
40-50us, the remaining ones can reach more than 80us.
We use \sys to analyze the max-segment and the min-segment and find that for all segments, the top causes are
NIC interrupts, but their impact values are different. We add the interrupt length
in Figure~\ref{fig:memcached-variance} and see a clear trend that segments with high
p99 latencies have high p99 interrupt lengths as well. More interestingly, we can see
that segments with high p99 latencies have lower p50 latencies and p50 interrupt lengths.
This indicates that, in the segments with high p99 latencies, interrupts are probably more bursty,
which causes a small number of requests to be affected by a large number of interrupts---this will
cause higher p99 latency but lower p50 latency.
The burstiness of the interrupts may be affected by network behavior,
which is beyond the scope of \sys, and we plan to investigate it in the future.

\begin{figure}[t]
        \centering
        \includegraphics[width=0.35\textwidth]{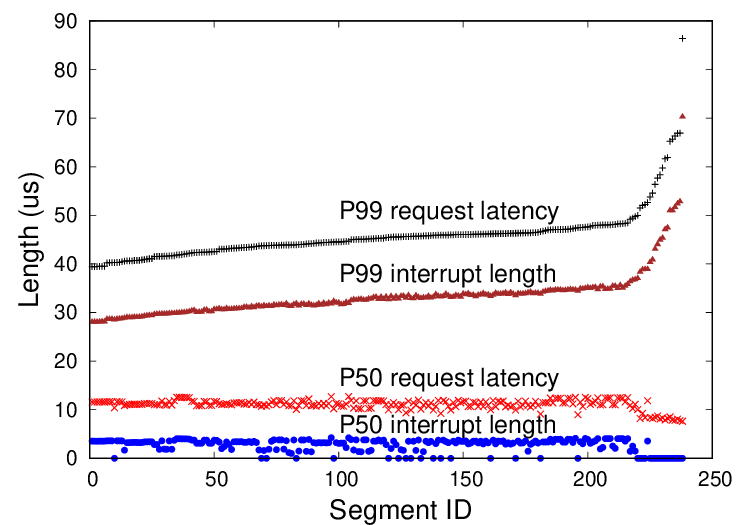}
        \caption{The request latencies and interrupt lengths of different segments in Memcached-get experiments (segments are sorted by their p99 request latency).}
         \vspace{-.15in}
        \label{fig:memcached-variance}
\end{figure}

For the set-only workload, \sys reports the top causes are scheduling events caused by lock contention
and kernel instructions related to \emph{sendmsg}.
The call stacks from \emph{perf} show that there are multiple
sources for the lock contention: 1) contention on a data item; 2) contention on the LRU list to determine which
data item to evict if the memory is full; and 3)
contention on the memory allocator. To verify this, we disable LRU update since our experiment
does not need to evict items, and this can reduce p99 of the median-segment by 31\% and
reduce CoV from 4.4\% to 1.2\%. A more practical optimization
would be to use multiple memory regions and assign an LRU list and a memory allocator for
each region.

We observe cross-segment variance for this test as well and \sys shows the reason
is the burstiness of lock contention.

\subsection{ZooKeeper}
ZooKeeper~\cite{Hunt10ZooKeeper} is 
a distributed lock service. ZooKeeper has a three-stage pipeline, with one thread in each stage: the first
stage sends and receives packets; the second stage pre-processes an incoming request;
the final stage logs the request to
disk if the request is a set operation, executes the request, and sends the reply to
the first stage. Therefore, we define three \utasks as the start and end of each stage
and we use T0, T1, and T2 to denote each \utask.

We populate ZooKeeper 3.4.11 with 10,000 10KB
items and test it with get-only and set-only workloads under 80\% maximal throughput.
We find T2 is much longer than others, so we
focus on T2.

For the get-only workload, \sys reports the top cause of variance is scheduling event that blocks T2, and the call stack
shows it is caused by waiting for the GC threads in JVM.
To confirm it, we reduce
GC by creating a buffer pool in ZooKeeper to re-use buffers of the same size.
Since page migration shows up in the report as well, we disable
NUMA balancing. These two optimizations reduce the p99 of the median-segment by 20\%
and reduce the CoV from 4\% to 1.9\%.

Our analysis on the cross-segment variance shows that it is highly related to the waiting for GC: in the max-segment, waiting for GC
has a much higher impact value; in the min-segment, waiting for GC is not the top cause.

For the set-only workload, \sys reports the top causes of tail latency in the median-segment are a
number of cache or instruction-related counters.
We don't find anything particularly interesting in their call stacks.

However, when investigating cross-segment variance, we find more interesting patterns: for both the median-segment
and the min-segment, the top causes are CPU counters; for max-segment, however, the top causes
include waiting for disk I/Os. We draw the I/O length of each segment in Figure~\ref{fig:zk-set-variance}: the p99 I/O lengths of a few segments are significantly higher than others, but their
p50 I/O lengths are not much different. It means these segments with high p99 latency are probably suffering from a disk slower at tail range.
We have confirmed this problem by running a disk benchmark on problematic and normal machines.

\begin{figure}[t]
        \centering
        \includegraphics[width=0.35\textwidth]{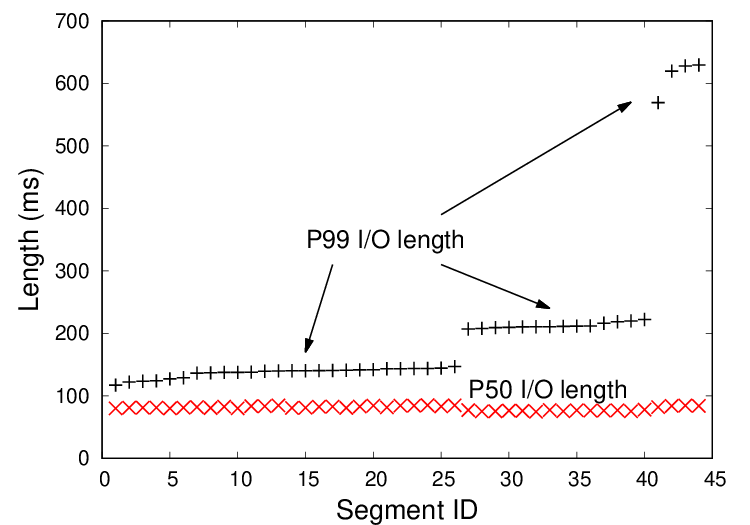}
        \caption{I/O lengths of different segments in ZooKeeper-set experiments. Request latency is dominated by I/O length so we only draw I/O length.}
        \vspace{-.2in}
        \label{fig:zk-set-variance}
\end{figure}

\subsection{MySQL}
 MySQL~\cite{mysql} is a popular relational database system.
MySQL creates a thread for each client to process its request (i.e., transaction), and since a request can
include multiple operations, we annotate the start and the end of executing an operation as an \utask.
We use the YCSB-T benchmark~\cite{dey2014ycsb+} to test MySQL 5.7.23. YCSB-T populates
the database with 10,000 rows and generates a mixed workload with 95\% selects and 5\% updates.
We use the default configuration
of YCSB-T that is zipfian distribution with 99\% skewness.
Since MySQL adds a commit operation and a setoption operation to each select or update request, we
annotate them as three \utasks, and we use T0 (select or update), T1 (commit), and T2 (setoption) to denote them.
We set MySQL to run on a mem-disk to prevent disk from becoming the bottleneck and test it under 80\%
maximal throughput. We find T0 is much longer, so we present the result of T0.

\sys reports scheduling events that switch threads to ``runnable'' is the
top reason of tail latency. This is because MySQL creates one thread for
each client and thus has many threads,
which causes frequent thread scheduling. Fixing this problem will need one thread to be able to process
requests from multiple clients and we observe a few other database systems and a plugin to
MySQL have already
adopted such a design~\cite{MySQLRef,Percona2010Limit,PostgreSQL,MariaDB}.

In addition, \sys reports several instruction counters, such
as the number of branch instructions, are the second cause of tail latency. 
Since many of their call stacks point to performance tracing of MySQL, we disable this feature.
\sys shows NUMA balancing is
another cause and we disable it as well.
These two combined can reduce the p99 of median-segment by 14\%
and reduce CoV from 5.6\% to 1.4\%.

Our analysis on the cross-segment variance shows that
it is related to the burstiness of the ``runnable'' scheduling.

\subsection{HBase} HBase~\cite{hbase} is a key-value store, which persists data to
HDFS~\cite{HdfsWebPage}. A client can send set and get requests to an HBase RegionServer
through RPCs, so we annotate the start and the end
of an RPC as an \utask. We test HBase with a set-only
workload and 1KB key-value pairs.
\sys (\emph{pTarget}=p99) reports the top cause is the waiting
for HDFS I/O to complete and we don't observe significant
cross-segment variance. Since the problem is outside of HBase, we do not further investigate.

\subsection{Lessons}

First, our experiments has re-confirmed the challenges of reproducing experiments
results: even with several hours of experiments, the tail latencies
of some applications are not stable, which will create
challenges for performance comparison~\cite{Maricq2018Variability}.
\sys's analysis may be able to help to alleviate this problem: if we
know some experiments are affected by abnormal factors like
slow disks, we can simply remove such experiments; 
for experiments affected by uncontrollable factors like bursty
interrupts, we may compare results under the same bustiness level.

Second, we find that \sys can provide useful hints but a more thorough investigation of the
root cause may require more specific tools, like cache behavior analysis~\cite{Pesterev2010Cache}, lock
contention analysis~\cite{Alam2017SyncPerf,Zhou2018wPerf}, heap analysis for GC~\cite{hprof,jmap}, etc, which
usually incur a higher overhead. 
This actually fits with our principle to record coarse-grained information first and record additional information
when necessary. Therefore, in practice, we expect 
these tools to be complementary: we can enable \sys first to narrow down the problem;
then we can target specific problems with specific tools.

\section{Evaluation}
\label{sec:overhead}

\vspace{.1in}
\noindent\textbf{Recording overhead.} We first present the detailed result on Memcached with get-only
workload since it has the highest event rate.
Figure~\ref{fig:memcached-overhead-throughput} shows the overhead of \sys with different selection rates.
This figure includes the overhead of \emph{ftrace}~\cite{ftrace} as a comparison (we add system calls
to record CPU counters in the same way as \sys).
As one can see, comparing \sys (100\%) to \emph{ftrace}, which records the timing of all events, \sys
can reduce trace size but will incur additional overhead to compute cumulative lengths. However, with \sys's support of selective tracing,
one can tune the selection rate to adjust overhead: 
with 1\% selection rate, \sys decreases the maximal throughput by about 0.49\% and generates about
1MB of trace per second.  We also test the overhead of \emph{perf}~\cite{perf}, which can record the call stacks of
 a subset of events.
We find its overhead is much larger: when recording 1\% of the events, \emph{perf} generates 24MB of trace per second.
We further compare the end-to-end latency of Memcached with and without \sys:
with 1\% selection rate, we do not see any observable difference.

The overhead on other applications varies, though recording all events is still
unacceptable for most of them in the long term: for Memcached-set, ZK-get, ZK-set, MySQL, and HBase, recording all events
can generate 101.3MB, 7.2MB, 0.9MB, 35.7MB, and 3.2MB of trace per second. Recall that we try to limit
trace rate to about 0.1MB/s in our studies.

\begin{figure}[t]
        \centering
        \includegraphics[width=0.4\textwidth]{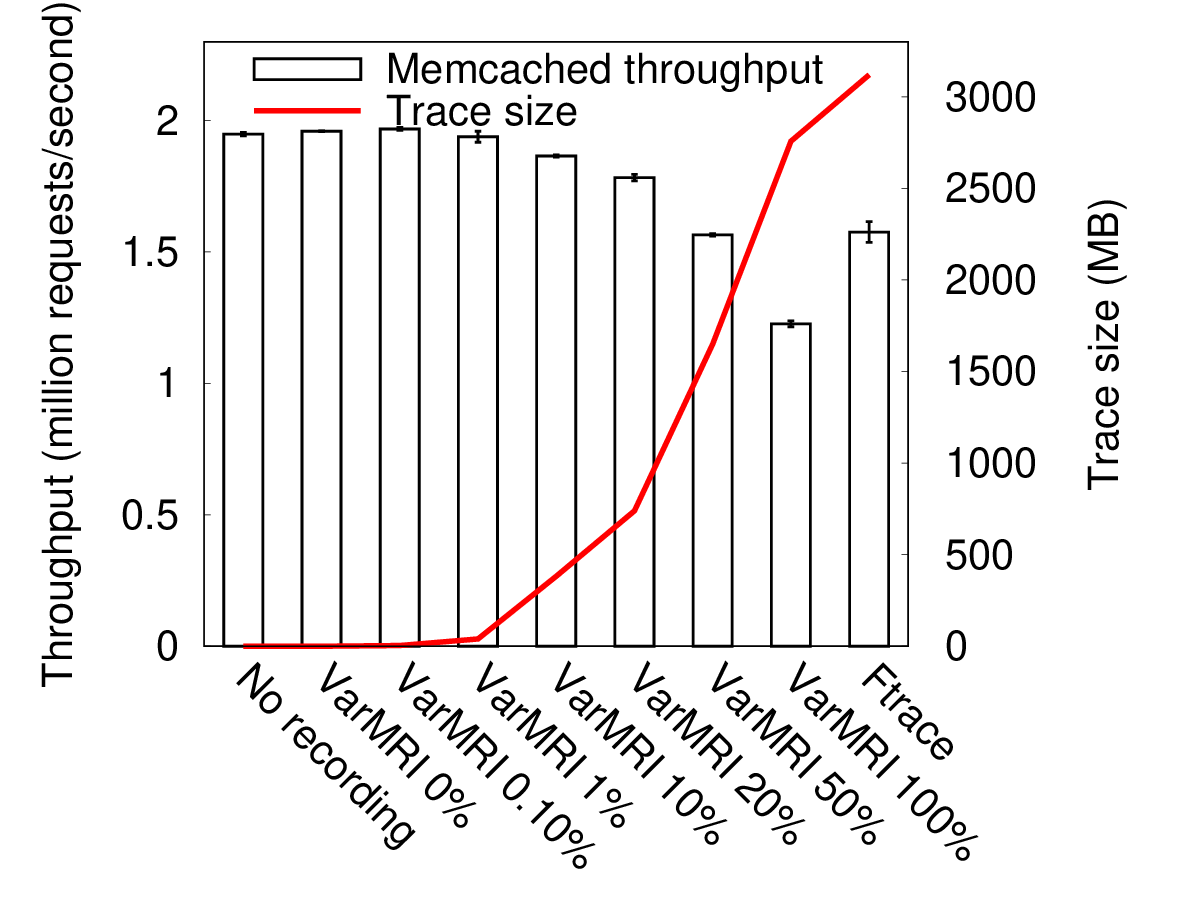}
        \vspace{-.1in}
        \caption{Overhead of applying \sys with different selection rates and \emph{ftrace} to Memcached (30s experiments).}
        \vspace{-.2in}
        \label{fig:memcached-overhead-throughput}
\end{figure}

\vspace{.1in}
\noindent\textbf{Analysis overhead.}
Analyzing one segment takes about 30-60 seconds in our experiments.
Considering the lengths of these segments (Table~\ref{table:tail-latency}), 
the analysis time is acceptable.

\vspace{.1in}
\noindent\textbf{Effects of domain-specific rules.}
We measure the number of events whose impacts would be among top 10
but are dropped out of top 10 because of the two domain-specific rules.
In total, the first rule filters 13 events and the second rule filters 14 events
for all the median segments we have analyzed (80 top-10 events in total).

	\section{Conclusion}
\label{sec:conclusion}

This paper builds \sys, a tool to support long-term
monitoring and analysis of kernel and hardware
events, and carries out a 3,000-hour study with the
help of \sys. The study reveals multiple factors
causing latency variance and how their impacts
can vary in the long term.
We plan to continue such studies on other
architecture and applications.

	\bibliographystyle{plain}
	\bibliography{LasrBibtex}

\end{document}